\documentclass[showpacs,preprintnumbers,amsmath,amssymb,superscriptaddress]{revtex4}

\usepackage{amsmath}
\usepackage{amssymb}
\usepackage{bm}
\usepackage{graphicx}
\usepackage{multirow}

\usepackage{textcomp}

\allowdisplaybreaks[1]

\begin{document}


\title{\boldmath Non-dipolar Wilson links for quasi-parton distribution functions}

\author{Hsiang-nan Li}
\affiliation{Institute of Physics, Academia Sinica,
Taipei, Taiwan 115, Republic of China}

\date{\today}

\begin{abstract}
We propose a modified definition for a quasi-parton distribution function (QPDF)
with an equal-time correlator in the large momentum limit, whose two pieces of
space-like Wilson links are oriented in orthogonal directions. It is explicitly
shown at one-loop level that the linear divergence in the original QPDF with dipolar
Wilson links, which complicates its matching to the standard light-cone
parton distribution function (LPDF), is removed. The LPDF can then be extracted
reliably from Euclidean lattice data for the QPDF with the non-dipolar Wilson links.

\end{abstract}


\maketitle

%
%
%

A quasi-parton distribution function (QPDF), which involves
an equal-time correlator in the large momentum limit with the Wilson links in the
direction $n=(0,0,0,1)$, has been proposed recently \cite{Ji:2013dva},
\begin{eqnarray}
{\tilde q}_n(x,\mu,P^z)=\int_{-\infty}^{\infty}\frac{dz}{4\pi}
e^{izk^z}\langle P|\psi(w)W^{\dagger}_{n}(w)\gamma^zW_{n}(0)\psi(0)|P\rangle.\label{qpdf1}
\end{eqnarray}
In the above expression $\gamma^z$ is a Dirac matrix, $\psi$ is the quark field,
$w=(0,0,0,z)$ is a coordinate, $|P\rangle$ is the nucleon state with the momentum
$P=(P^0, 0,0, P^z)$, $x=k^z/P^z$ is the momentum fraction of the quark,
the large scale $\mu$ represents a ultraviolet cutoff for loop momenta,
the gauge link is given by
\begin{eqnarray}
W_{n}(w) =P \exp \left[ -i g\int_0^{\infty} \, d \lambda \, T^a \,
n \cdot A^a (\lambda n +w) \right],
\end{eqnarray}
with a color matrix $T^a$, and the average over the nucleon spin is implicit. The
definition in Eq.~(\ref{qpdf1}) was motivated by a direct evaluation of a parton density
on a Euclidean lattice through Lorentz boost. Note that the standard light-cone parton
distribution function (LPDF), based on a time-dependent correlation with
the two quark fields being located on a light cone, cannot be directly computed.
Instead, one has to work on moments of the LPDF, which are matrix elements of
local operators from the expansion of the nonlocal correlator. However, it has
been known that the analysis becomes technically challenging for higher moments
\cite{JWN}.

It has been shown that the QPDF contains the same collinear logarithmic
divergence as in the LPDF to all orders in the strong coupling constant $\alpha_s$
\cite{Ma:2014jla}. In principle, the LPDF $q$ is
extracted from Euclidean lattice data for ${\tilde q}_n$
\cite{Lin:2014zya, Alexandrou:2015rja} via the matching formula
\begin{eqnarray}
{\tilde q}_n(x,\mu,P^z)=\int \frac{dy}{y}Z\left(\frac{x}{y},\frac{\mu}{P^z}\right)
q(y,\mu),\label{fac}
\end{eqnarray}
where $Z$ is an infrared finite kernel.
The strategy of deriving a LPDF from lattice data for a correlator of two operators
with a space-like separation can be traced back to \cite{Aglietti:1998ur,
Abada:2001if,Braun:2007wv}. Their idea is to calculate a lattice ``pion transition form
factor" with the two electromagnetic currents being separated by a space-like distance,
from which the pion light-cone distribution amplitude is extracted. The proposal
of \cite{Ma:2014jla} to obtain the LPDF from lattice ``cross sections" is similar,
in which the QPDF is regarded as a special case of lattice ``cross sections".
Though a direct evaluation of a parton density becomes feasible,
Eq.~(\ref{qpdf1}) suffers a linear power divergence proportional to $\mu/P^z$ from
radiative gluons attaching to the Wilson links.
Whether the QPDF can be factorized into the convolution in Eq.~(\ref{fac})
at higher-order accuracy for $Z$ in the presence of the linear
divergence is not certain \cite{Ma:2014jla}. We will demonstrate that
Eq.~(\ref{fac}) breaks down at two-loop level actually
due to an additional collinear divergence induced by the linear divergence.
This additional collinear divergence, going into $Z$ after the matching, fails
perturbative expansion for $Z$, and leads to uncontrollable theoretical uncertainty
in the extraction. It should be mentioned that the difference between the QPDF and
the LPDF for light quarks has been investigated in the framework of the spectator
diquark model, and found to be about 20-30\% for $P^z$ around few GeV
\cite{Gamberg:2014zwa}.

It is interesting to notice that a linear pinched singularity
appears in a transverse-momentum-dependent (TMD) parton density
with dipolar Wilson links off the light cone \cite{BBDM}. Actually,
the diagrams contributing to the linear divergences in the QPDF and in a
TMD parton density, i.e., those from real gluon exchange
between the two pieces of Wilson links, are identical. We have demonstrated
\cite{Li:2014xda} that the linear pinched singularity can be removed by introducing
non-dipolar Wilson links, namely, Wilson links each of which is oriented in different
directions, and that the collinear divergence to be factorized into a TMD
parton density is maintained. The simplest version is the one with
the two pieces of Wilson links being orthogonal to
each other. Motivated by the above observation, we
propose an improved definition for the quark QPDF without the linear divergence,
\begin{eqnarray}
{\tilde q}(x,\mu,P^z)=\int_{-\infty}^{\infty}\frac{dz}{4\pi}
e^{izk^z}\langle P|\psi(w)W^{\dagger}_{n_2}(w)\gamma^zW_{n_1}(0)\psi(0)|P\rangle,\label{mod}
\end{eqnarray}
where $n_1=(0,1,0,1)$ and $n_2=(0,-1,0,1)$ denote the directions of the two orthogonal
pieces of non-dipolar Wilson links. A vertical link in the $x$-direction, which
connects the two pieces of non-dipolar Wilson links at infinity, is understood.
Such a link at infinity does not contribute in covariant gauge.
A pair of non-dipolar Wilson links can be well approximated by zig-zag gauge links
on lattice.

In this Letter we will show up to one-loop level that the modified QPDF in
Eq.~(\ref{mod}) does not develop the linear divergence, and exhibits the
infrared behavior the same as of the LPDF. It will be explained that the
factorization formula in Eq.~(\ref{fac}) holds for the modified QPDF to all orders,
but breaks down for the original QPDF in Eq.~(\ref{qpdf1}) at two loops, whose
kernel $Z$ gains infrared divergences at this
level of accuracy. The modified QPDF is then matched to the LPDF
with an infrared finite kernel $Z$ calculable in a perturbation theory, so that the
latter can be extracted reliably from lattice data for the former.
Equation~(\ref{mod}) yields the result the same as the quark LPDF at leading order,
\begin{eqnarray}
\tilde q^{(0)}(x)=q^{(0)}(x)=\delta(1-x).
\end{eqnarray}
For next-to-leading-order calculation, we introduce a gluon mass $m_g$ to regularize
infrared divergences, and a cutoff $\mu$ to regularize ultraviolet divergences.

\begin{widetext}
\begin{figure*}
 \centering
\includegraphics[scale=0.80]{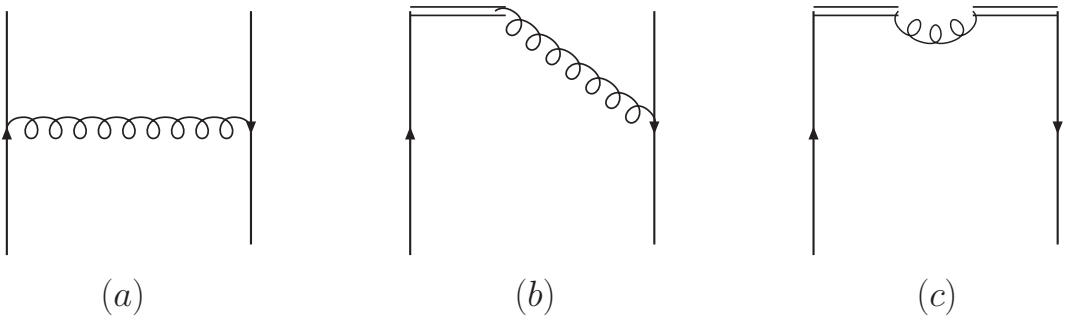}
\caption{\label{pdfr}
One-loop real corrections to the quasi-parton distribution function.}
\end{figure*}
\end{widetext}

The real gluon exchange between the two quarks in Fig.~\ref{pdfr}(a) gives
\begin{eqnarray}
\tilde q^{1a}&=&\left\{ \begin{array}{ll}
\frac{\alpha_s}{2\pi}C_F(1-x)\ln\frac{2(1-x)P^z
[\Lambda-(1-x)P^z]}{|x|m_g^2}, & \mbox{for $x<1$};\\
\frac{\alpha_s}{2\pi}C_F(1-x)\ln\frac{\Lambda-(1-x)P^z}{2(x-1)P^z},
& \mbox{for $x>1$},\end{array} \right.,\label{pdfra}\\
q^{1a}&=&\left\{ \begin{array}{ll}
\frac{\alpha_s}{2\pi}C_F(1-x)\ln\frac{\mu^2}{|x|m_g^2},& \mbox{for $x<1$};\\
0,& \mbox{for $x>1$},\end{array} \right.,\label{lpdfa}
\end{eqnarray}
for the modified QPDF and the LPDF, respectively, with the cutoff
$\Lambda=\sqrt{\mu^2+[(1-x)P^z]^2}\approx \mu$. It is seen that the modified QPDF
does not produce a collinear divergence denoted by $\ln m_g^2$ for $x>1$ as expected,
since the LPDF does not either. For $x<0$, the quark lines in Fig.~\ref{pdfr}(a)
can go on-shell \cite{Collins:2011zzd}, contributing collinear divergences which
cancel those in Eqs.~(\ref{pdfra}) and (\ref{lpdfa}).
That is, the modified QPDF and the LPDF develop the same collinear divergence
only in the physical support region $0<x\le 1$ actually. Below we consider the QPDF
result for $0<x<2$, namely, $-P^z<l^z<P^z$. It will be observed that the QPDF may
contain a soft divergence from $x\to 1$ in the range $x>1$, which should be taken
into account as determining the matching in the range $0<x\le 1$.

For the LPDF with the light-like Wilson links in the direction $n_-=(1,0,0,-1)$,
Fig.~\ref{pdfr}(b) generates a logarithmic divergence,
\begin{eqnarray}
q^{1b}&=&g^2C_F \int_0^{\mu}\frac{d^{2}l_T}{(2\pi)^{3}}
\frac{P^+(P^+-l^+)}{l^+[P^+l_T^2+(P^+-l^+)m_g^2]}\nonumber\\
&=&\frac{\alpha_s}{4\pi}C_F\frac{2x}{1-x}\ln\frac{\mu^2}{xm_g^2},
\label{srb}
\end{eqnarray}
with the plus component $l^+=(1-x)P^+$ and the transverse component
$l_T=\sqrt{(l^x)^2+(l^y)^2}$ of the loop momentum.
The loop integral for the modified QPDF corresponding to Fig.~\ref{pdfr}(b)
is written as
\begin{eqnarray}
\tilde q^{1b}&=&-\frac{1}{4}g^2C_F\int\frac{d^{4}l}{(2\pi)^{3}}
{\rm tr}\left[\not P\gamma_\nu\frac{\not P-\not l}{(P-l)^2}\gamma^z\right]
\frac{n_1^\nu}{n_1\cdot l}\delta(l^2-m_g^2)\delta(P^z-k^z-l^z)\nonumber\\
&=&\frac{1}{2}g^2C_F\int\frac{d^{2}l_T}{(2\pi)^{3}}
\frac{P^z(2P^z-l^0-l^x-l^z)}{l^0(l^x+l^z)[2P^z(l^0-l^z)-m_g^2]},\label{pdfrb}
\end{eqnarray}
with the components $l^0=\sqrt{l_T^2+(l^z)^2+m_g^2}$ and $l^z=(1-x)P^z$.
The rotation of the Wilson link off the $z$ axis
changes the eikonal propagator from $1/l^z$ to $1/(l^x+l^z)$.
The scaling of the modified eikonal propagator $1/(l^x+l^z)\sim 1/l_T$
at large $l_T$ then renders the integrand in Eq.~(\ref{pdfrb}) scale like $1/l_T^2$,
similar to that in Eq.~(\ref{srb}).

Working out the integration over $l_T$, we obtain
\begin{eqnarray}
\tilde q^{1b}=\left\{ \begin{array}{ll}
\frac{\alpha_s}{4\pi}C_F\left[
\frac{2x}{1-x}\ln\frac{2(1-x)P^z\mu}{x m_g^2}
-\frac{1+x}{1-x}\ln\frac{\mu}{(1-x)\sqrt{2}P^z}
-\frac{\pi}{4}\right],& \mbox{for $0<x<1$};\\
\frac{\alpha_s}{4\pi}C_F\left[\ln\frac{\mu}{2(x-1)P^z}
+\frac{1+x}{1-x}\ln\sqrt{2}+\frac{\pi}{4}\right],& \mbox{for $x>1$}.\end{array}
\right..\label{pdfrb1}
\end{eqnarray}
No collinear divergence, but a soft divergence is produced in the limit $x\to 1$
from the region of $x> 1$. It is confirmed that the LPDF and
the modified QPDF from Fig.~\ref{pdfr}(b) involve the identical collinear logarithm
$\ln m_g^2$ for $0<x < 1$. The real vertex correction associated with another
Wilson link in the direction of $n_2$ can be derived by flipping the sign of the
component $l^x$ in Eq.~(\ref{pdfrb}). Apparently, the result is equal to
$\tilde q^{1b}$ in Eq.~(\ref{pdfrb1}).

We explain the origin of the linear divergence in the original QPDF from
Fig.~\ref{pdfr}(c), where the real gluon is exchanged between the two Wilson links
in the $z$ direction:
\begin{eqnarray}
\tilde q^{1c}_n&=&-\frac{1}{2}g^2C_F\int \frac{d^2l_T}{(2\pi)^{3}}
\frac{P^z}{l^0(l^z)^2},\nonumber\\
&=&-\frac{\alpha_s}{2\pi}C_F
\frac{\mu}{(1-x)^2P^z}.\label{pdfrc}
\end{eqnarray}
It is noticed that the denominator scales like $l^0\sim l_T$,
because the component $l^z$ is specified to be $(1-x)P^z$. According to \cite{XJZZ},
the linear divergence comes only from Fig.~\ref{pdfr}(c) for the original QPDF.
The above linear divergence has an origin similar to that of the known light-cone
singularity \cite{Co03} appearing in the naive definition of a TMD parton density.
The latter arises from the region with a loop momentum parallel to a Wilson
link along, say, $n_-$, namely, the region with $v\cdot l \gg n_-\cdot l$,
where $v^\mu\equiv P^\mu/P^z$ represents the nucleon velocity. The former arises
from the region with $v\cdot l\gg |n\cdot l|$, namely, $l^0\gg l^z$.

As elaborated below, the linear divergence induces an additional collinear
divergence at two-loop level from the region with loop momenta
being perpendicular to the Wilson link, i.e., to the $z$ direction.
The contribution from this special region is usually of subleading power,
but a subleading-power collinear divergence in the case of the original
QPDF, due to the presence of the linear divergence, is not power-suppressed.
The additional collinear divergence, characterized by $l^0\sim l_T\gg l^z$,
differs from the ordinary collinear divergence in LPDF, characterized by
$l^0\sim l^z \gg l_T$, and cannot be absorbed into the LPDF. That is,
it renders the matching kernel $Z$ infrared divergent, and breaks the
factorization in Eq.~(\ref{fac}), such that the LPDF cannot be extracted
reliably from the lattice data of the original QPDF. With the additional
collinear divergence, which does not exist in QCD diagrams for a considered
process, the original QPDF is not compatible with a factorization theorem
either: the corresponding hard kernel, defined as the difference between QCD
diagrams and the original QPDF, then contains the additional collinear
divergence, and is not calculable in a perturbation theory.

We will verify the above observation at two loops below by working in the
light-cone gauge $n_-\cdot A=0$, in which the original collinear
divergence is killed, and any survived collinear divergence cannot be absorbed
into the LPDF. Since we focus on the additional collinear divergence
induced by the linear divergence, we need to consider only the diagram with
the second gluon attaching to a quark line and to the gluon line in
Fig.~\ref{pdfr}(c). The on-shellness of the two gluon propagators,
which the first gluon momentum $l_1$ flows through, leads to the additional
collinear divergence from $l_1^0\sim l_{1T}\gg l_1^z$. The
diagrams where the second gluon attaches to a quark line and a Wilson line
do not generate a collinear divergence in the light-cone gauge. Those
with the second gluon attaching to other one-loop diagrams do not either
due to the absence of the linear divergence, i.e., to the power suppression
on the region with $l_1^0\sim l_{1T}\gg l_1^z$. Hence, it is impossible for the
identified additional collinear divergence
to be cancelled by other two-loop diagrams in the light-cone gauge.

Let the second gluon carry the momentum $l_2$, and label the
vertex on the quark line after the final-state by $\mu$, and the
vertex on the Wilson link before (after) the final-state cut by
$\nu$ ($\lambda$). The fermion trace for the considered Feynman diagram
is proportional to $tr[\not P\gamma^\mu(\not P+\not l_2)\gamma^+]
\propto P^\mu(P^++l_2^+)+(P^\mu+l_2^\mu)P^+$,
where the projector $\gamma^z$ has been replaced by $\gamma^+$ because of
$\not P=P^+\gamma^-$. The gluon propagator in the light-cone gauge satisfies
\begin{eqnarray}
l_{2\mu} d^{\mu\mu'}(l_2)&=&l_{2\mu} \left(-g^{\mu\mu'}+\frac{n_-^\mu l_2^{\mu'}
+n_-^{\mu'} l_2^\mu}{n_-\cdot l_2}\right)\to 0,
\label{ide1}
\end{eqnarray}
for an almost real gluon with $l_2^2\to 0$,
so we evaluate only the contribution from $P^\mu$, i.e., from
$v^\mu$. The two-loop diagram contains the partial integral
\begin{eqnarray}
I&=&\int^\mu\frac{d^2 l_{1T}}{l_1^0}
\frac{\Gamma(l_1^0,l_1^z;l_2^0,l_2^z)}{(l_1+l_2)^2},\label{int}
\end{eqnarray}
where the factor
\begin{eqnarray}
\Gamma&=&[(2l_1+l_2)^{\mu'} g^{\nu'\lambda'}+(l_2-l_1)^{\lambda'} g^{\mu'\nu'}
-(l_1+2l_2)^{\nu'} g^{\mu'\lambda'}]d_{\mu\mu'}(l_2)d_{\nu\nu'}(l_1)
d_{\lambda\lambda'}(l_1+l_2)v^\mu n^\nu n^\lambda\nonumber\\
&=&\frac{v\cdot(l_1+2l_2)n\cdot l_1}{n_-\cdot l_1}
-\frac{v\cdot(2l_1+l_2)n\cdot l_2}{n_-\cdot l_2}
+\frac{v\cdot(l_1-l_2)n\cdot (l_1+l_2)}{n_-\cdot (l_1+l_2)},\label{it}
\end{eqnarray}
comes from the product of the triple-gluon vertex and the gluon propagators in
the light-cone gauge, $n^2=-1$ has been inserted, and terms proportional to
the small invariants $l_1^2$, $l_2^2$, and $l_1\cdot l_2$ have been dropped.
The denominator $l_1^0=\sqrt{l_{1T}^2+(l_1^z)^2+m_g^2}$ is a
result of the integration of $\delta(l_1^2-m_g^2)$ or $1/(l_1^2-m_g^2)$ over
the loop component $l_1^0$. It suffices to regularize a collinear divergence
by associating $m_g^2$ only with $l_1^2$.

For the second gluon moving in the perpendicular direction, we have a
typical configuration with $l_2^0\sim l_{2T}\sim O(P^z)\gg l_2^z$.
It is straightforward to extract the collinear divergence in Eq.~(\ref{int}),
\begin{eqnarray}
I&=&\frac{\pi l_1^zl_2^2}{(l_2^z)^3}
\ln\frac{l_1^z(l_2^0)^2m_g^2}{4\mu^2(l_2^z)^3}+\cdots,\label{int1}
\end{eqnarray}
where the companying linear divergence is not shown explicitly. The above collinear
divergence occurs as $l_1^0\approx l_1^zl_2^0/l_2^z\gg l_1^z\sim O(P^z)$, namely,
as the two gluons are collimated to each other in the perpendicular direction. The
factor $l_2^2$ in Eq.~(\ref{int1}) cancels the propagator of the second gluon,
ensuring that the integration over $l_2$ is infrared finite as expected, and of
order unity. That is, the additional collinear divergence from the special region
is not power suppressed. Because $q^{(1)}$ and $q^{(2)}$ have been known from the
LPDF, and $Z^{(1)}$ has been defined in the one-loop analysis, the divergence in
Eq.~(\ref{int1}) can only go into the convolution $Z^{(2)}\otimes q^{(0)}$, rendering $Z$
infrared divergent at two loops. Therefore, the factorization formula in Eq.~(\ref{fac}),
which is supposed to define an infrared finite $Z$, breaks down at higher orders
in $\alpha_s$ for the original QPDF. It is easy to confirm that the special region,
without the linear divergence, is power suppressed in the other two-loop diagrams
for the original QPDF and in the case of the modified QPDF .

The contribution from Fig.~\ref{pdfr}(c) vanishes for the LPDF, since its
integrand is proportional to $n_-^2=0$.
The linear divergence is removed by our choice of the orthogonal
Wilson links with $n_1\cdot n_2=0$ for the modified QPDF, namely,
$\tilde q^{1c}=0$. We stress that even if $n_1\cdot n_2\not=0$, the rotation of
the Wilson links away from the $z$ direction will still suppress the linear
divergence: the modified eikonal propagator, depending on $l^x$, scales like $1/l_T$
at large $l_T$. The corresponding integrand [referred to Eq.~(\ref{pdfrc})]
then scales like $1/l_T^3$, and the integral over $l_T$ is convergent.
The observation is that the boost direction of the nucleon should differ from the
direction of the associated Wilson links in order to avoid the linear divergence.
For instance, a nucleon is boosted in the plus
direction, while the Wilson links run in the minus direction in the LPDF case.

\begin{widetext}
\begin{figure*}
 \centering
\includegraphics[scale=0.80]{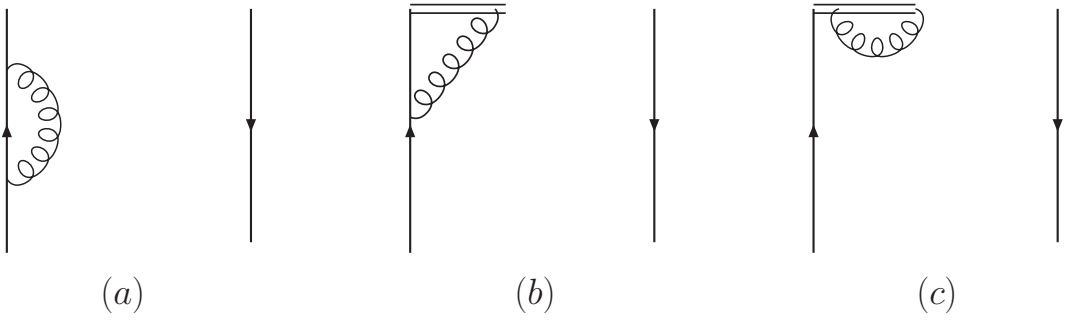}
\caption{\label{pdfv}
One-loop virtual corrections to the quasi-parton distribution function.}
\end{figure*}
\end{widetext}

The evaluation of the self-energy correction to a quark field in Fig.~\ref{pdfv}(a)
is trivial, and the expression
\begin{eqnarray}
\tilde q^{2a}&=&-\frac{\alpha_s}{4\pi}C_F\int_0^1dy
y\ln\frac{\mu^2}{y m_g^2}\delta(1-x),\label{pdfva}
\end{eqnarray}
is the same as for the LPDF and for the original QPDF.

The loop integration for the virtual quark-Wilson-link vertex correction in
Fig.~\ref{pdfv}(b),
\begin{eqnarray}
\tilde q^{2b}&=&-ig^2C_F\int\frac{d^{4}l}{(2\pi)^{4}}
\frac{P^z(l^0+l^x+l^z+2P^z)\delta(P^z-k^z)}{(l^x+l^z-i\epsilon)
(l^2+2P\cdot l+i\epsilon)(l^2-m_g^2+i\epsilon)},
\label{pdfvb}
\end{eqnarray}
is more difficult to carry out. We work in the light-cone coordinates, in which
contributions from both $l^z>0$ and $l^z<0$ can be included simultaneously.
For $-P^+<l^+<0$ ($l^+<-P^+$), we pick up the pole
$l^-=(l_T^2+m_g^2)/(2l^+)+i\epsilon$ ($l^-=l^++\sqrt{2}l^x-i\epsilon$)
from the gluon (eikonal) propagator. Equation~(\ref{pdfvb}) gives
\begin{eqnarray}
\tilde q^{2b}&=&\frac{\alpha_s}{4\pi}C_F\int_{0}^{1}dy
\left\{\frac{1+y}{1-y}\ln\frac{\mu}{(1-y)P^z}
-\frac{2y}{1-y}\ln\frac{2(1-y)P^z\mu}{ym_g^2}\right.\nonumber\\
& &\hspace{2.0cm}\left.+\frac{y}{1-y}\ln y-\frac{1}{1-y}\ln\frac{1+y}{2}
+\ln[\sqrt{2}(1-y)]\right\}\delta(1-x),\label{vb2}
\end{eqnarray}
where only the first line yields a soft divergence from $y\to 1$. For a similar
reason, the virtual vertex correction associated with another Wilson link
in the direction of $n_2$ is equal to $\tilde q^{2b}$ in
Eq.~(\ref{vb2}).

Figure~\ref{pdfv}(b) leads to
\begin{eqnarray}
q^{2b}=-\frac{\alpha_s}{4\pi}C_F\int_{0}^1dy\frac{2y}{1-y}
\ln\frac{\mu^2}{y m_g^2}\delta(1-x),
\end{eqnarray}
for the LPDF. It is seen that
the soft divergences from $x\to 1$ in $q^{1b}$ and from $y\to 1$ in
$q^{2b}$ cancel each other as expected, after a plus function is introduced.
Similar soft cancellation also takes place between $\tilde q^{1b}$ and $\tilde q^{2b}$,
once the soft divergences in the former from both the ranges $0<x< 1$ and $1<x<2$
are combined into a term proportional to $\delta(1-x)$ \cite{JSXY}. As a consequence,
only the collinear divergences represented by the logarithm $\ln m_g^2$ remain in
the sum of the real and virtual corrections.

At last, the self-energy correction to the Wilson link in Fig.~\ref{pdfv}(c)
should be subtracted from the definition of a parton distribution function, because
it is not involved in the Ward identity for the factorization of a parton
distribution function \cite{Collins:2011zzd}. As to an explicit gauge-invariant
subtraction of the Wilson link self-energy corrections, one may, according to
\cite{Collins:2011zzd}, simply introduce a denominator into the modified QPDF definition, 
which is a vacuum matrix element of three pieces of Wilson links: the first piece runs
from the origin along the direction $n_1=(0,1,0,1)$ to infinity, the second piece is
a vertical line in the $x$-direction located at infinity, and the third piece runs
from infinity along $n_2=(0,-1,0,1)$ back to the origin. These three pieces form a
close loop, and produce the Wilson link self-energy corrections the same as in the
modified QPDF, so this vacuum matrix element serves as a gauge-invariant subtraction.
Since Fig.~\ref{pdfv}(c) does not contribute to the kernel $Z$, we will not present
its explicit expression here. Note that the subtraction of the self-energy correction, 
if applied to the original QPDF, cannot restore the factorization formula in 
Eq.~(\ref{fac}), because the two-loop diagram discussed before is not removed by
the subtraction.

It is verified that the modified QPDF with the non-dipolar Wilson links
exhibits the same infrared behavior as of the LPDF at one-loop level:
\begin{eqnarray}
q^{(1)}&=&\frac{\alpha_s}{2\pi}C_F\left[\frac{1+x^2}{(1-x)_+}\ln\frac{\mu^2}{xm_g^2}
+\left(\frac{3}{2}\ln\frac{\mu^2}{m_g^2}+\frac{7}{4}-\frac{\pi^2}{3}\right)\delta(1-x)\right],
\label{q1}\\
\tilde q^{(1)}&=&\frac{\alpha_s}{2\pi}C_F\left\{
\frac{1+x^2}{(1-x)_+}\ln\frac{2P^z\mu}{xm_g^2}
+\frac{1}{2}\frac{1+x}{(1-x)_+}\ln\frac{2(P^z)^2}{\mu^2}+(2+x+x^2)\left[\frac{\ln(1-x)}{1-x}\right]_+
-\frac{\pi}{4}\right.\nonumber\\
& &\hspace{1.0 cm}\left.+\left[\frac{3}{2}\ln\frac{\mu^2}{m_g^2}+3\ln\frac{2P^z}{\mu}
-\frac{5}{4}-\frac{1}{2}(\ln^2 2+\ln 2)-\frac{5}{12}\pi^2\right]\delta(1-x)\right\},\label{tq1}
\end{eqnarray}
for $0<x\le 1$, despite of their different ultraviolet structures.
We then derive the infrared finite kernel up to $O(\alpha_s)$,
as matching Eq.~(\ref{tq1}) to Eq.~(\ref{q1}),
\begin{eqnarray}
Z\left(\xi,\frac{\mu}{P^z}\right)&=&\left[1-\frac{\alpha_s}{4\pi}C_F
\left(6\ln\frac{\mu}{2P^z}+6+\ln^2 2+\ln 2+\frac{\pi^2}{6}\right)\right]\delta(1-\xi)\nonumber\\
& &+\frac{\alpha_s}{4\pi}C_F\left\{
2(2+\xi+\xi^2)\left[\frac{\ln(2P^z/\mu)}{(1-\xi)_+}+\left(\frac{\ln(1-\xi)}{1-\xi}\right)_+\right]
-\frac{1+\xi}{(1-\xi)_+}\ln 2-\frac{\pi}{2}\right\},\label{z}
\end{eqnarray}
for $0<\xi\le 1$, whose logarithm $\ln(\mu/P^z)$ can be organized to all orders by the
standard renormalization-group method. Since we have also obtained the one-loop corrections
to the modified QPDF in the region with $x>1$, it is trivial to get the
corresponding infrared finite kernel
\begin{eqnarray}
Z\left(\xi,\frac{\mu}{P^z}\right)&=&\frac{\alpha_s}{4\pi}C_F\left[
2(\xi-2)\ln\frac{2(\xi-1)P^z}{\mu}-\frac{\xi+1}{(\xi-1)_+}\ln 2+\frac{\pi}{2}\right],\label{z1}
\end{eqnarray}
for $\xi>1$. The plus function in the above expression means that the soft divergence
in the integral $\int_1^2 d\xi/(\xi-1)$ is subtracted, as the kernel is convoluted with
other functions. Certainly, the matching must be recalculated up to a constant accuracy
using the standard approach on lattice \cite{SC}. The contribution related to the vertical
link in the $x$ direction, located at finite $z$ in this case, must be taken into account.

We postulate that the modified QPDF, which contains only the logarithmic collinear
divergence, respects the factorization formula in Eq.~(\ref{fac}) to all orders in $\alpha_s$.
The proof follows the procedures outlined in \cite{Li:2000hh}, including the
eikonal approximation for collinear gluons, the Ward identity applied to the summation
over all collinear gluon attachments, and the induction to extend the
factorization of the LPDF from lower orders to higher orders. Hence, once the proof was
done at one loop without the linear divergence in this Letter, the subleading-power
collinear divergence is power suppressed and neglected at higher loops, and the
factorization can be generalized to all orders straightforwardly.
The detail will be presented in a forthcoming paper. It is then justified to extend the
construction of $Z$ from the modified QPDF to higher orders as argued in \cite{Ma:2014jla}.

To summarize, our analysis has indicated that the regularization of
the linear divergence cannot sustain the factorization property of
the original QPDF at two loops. The modified QPDF with the non-dipolar Wilson
links is free of the linear divergence, and generates the same collinear logarithm
as in the LPDF at one-loop level, albeit with a distinct ultraviolet structure.
The modified QPDF does not only facilitate a factorization theorem, but allows a
reliable extraction of the LPDF from its Euclidean lattice data through perturbative
matching. The non-dipolar Wilson links proposed in this paper
can be applied to other nonperturbative objects, such as the gluon QPDF, polarized parton
distribution functions \cite{XJZZ}, generalized parton
distributions and TMD parton densities \cite{JSXY}.
These subjects will be explored in future works.

\begin{acknowledgments}


We thank J.W. Chen, H.Y. Cheng, X. Ji, D. Lin, Y.Q. Ma, J.W. Qiu, Y.M. Wang,
J.H. Zhang for illuminating discussions. This work was supported  in part by
the Ministry of Science and Technology of R.O.C. under
Grant No. MOST-104-2112-M-001-037-MY3.
\end{acknowledgments}

\end{document}